\documentclass[preprint,aps,superscriptaddress,showpacs]{revtex4}
\usepackage{graphicx}
\begin{document}

\title{Staggered Spin Order of Localized $\pi$-electrons in the
Insulating State of the Organic Conductor $\kappa$-(BETS)$_2$Mn[N(CN)$_2$]$_3$}

\author{O.~M.~Vyaselev}
\affiliation{Institute of Solid State Physics, Russian Academy of Sciences,
Chernogolovka, Moscow region, 142432, Russia}

\author{M.~V.~Kartsovnik}
\affiliation{Walther-Meissner-Institut, Bayerische Akademie der Wissenschaften,
Garching, Germany}

\author{N.~D.~Kushch}\author{E.~B.~Yagubskii}
\affiliation{Institute of Problems of Chemical Physics, Russian Academy of Sciences,
Chernogolovka, Moscow region, 142432, Russia}

\begin{abstract}

Magnetic properties of the conduction $\pi$-electron system of
$\kappa$-(BETS)$_2$Mn[N(CN)$_2$]$_3$ have been probed using $^{13}$C NMR. At
ambient pressure, the metal-insulator transition observed in the resistivity
measurements below $T\simeq23\,$K is shown to be accompanied by ordering of the
$\pi$-spins in a long-range staggered structure. As the metal-insulator
transition is suppressed by applying a small pressure of $\sim 0.5\,$kbar, the
$\pi$ spin system maintains the properties of the metallic state down to 5\,K.

\end{abstract}


\maketitle

\begin{figure}[p]
\includegraphics[width=0.75\linewidth,clip]{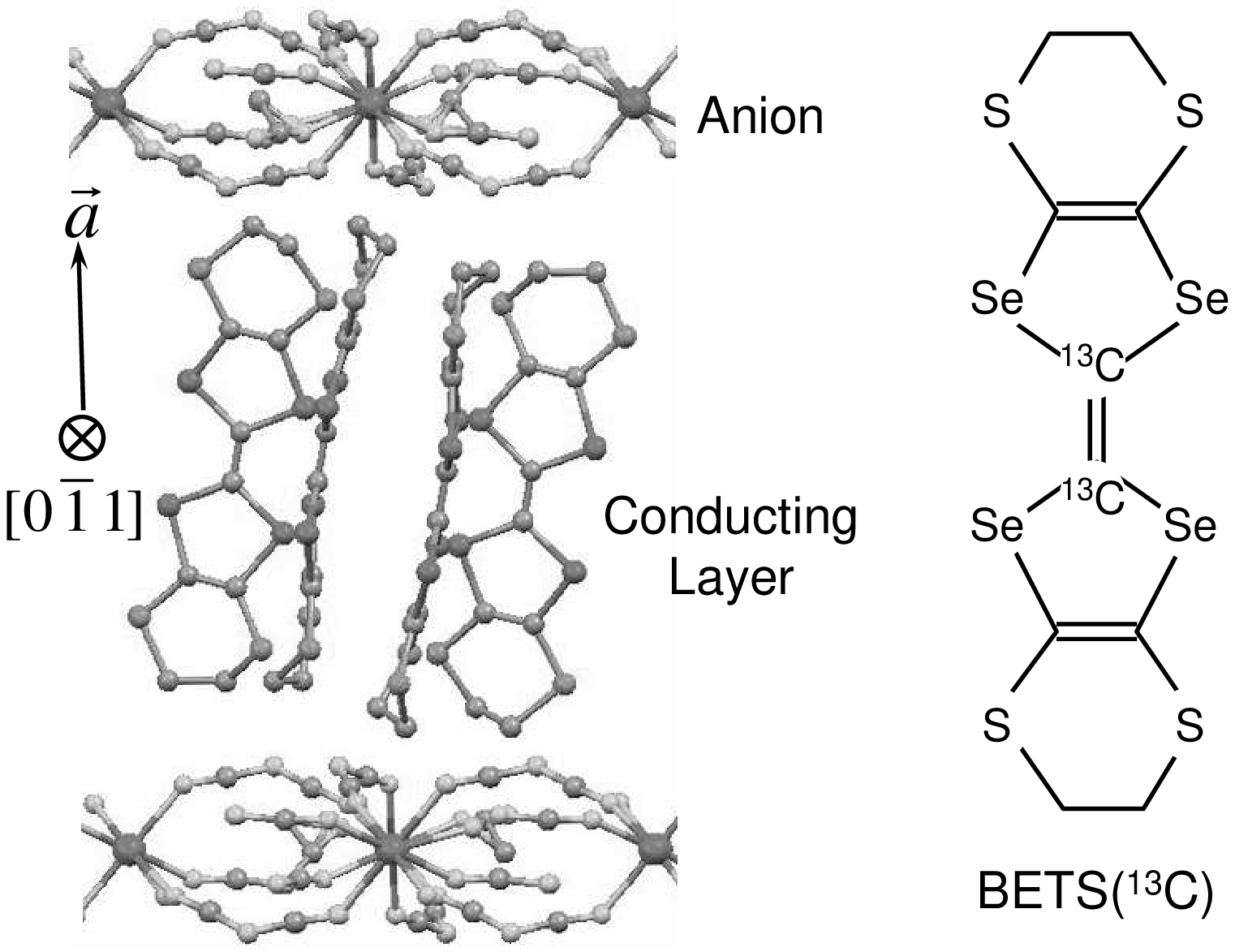}
\caption{The crystal structure of BETS-Mn viewed along $[0\bar{1}1]$ direction
(left), and the BETS($^{13}$C) molecule with $^{13}$C isotope in the positions
of the central carbon sites (right). \label{figStruct}}
\end{figure}

The organic charge-transfer salt $\kappa$-(BETS)$_2$Mn[N(CN)$_2$]$_3$
\cite{r12} (hereafter abbreviated as BETS-Mn) is a quasi-two-dimensional metal
consisting of insulating layers of Mn[N(CN)$_2$]$_3^-$ polymer anions
alternating with conducting layers of dimers of organic radical cations BETS
[bis(ethylenedithio)tetra-selenafulvalene, C$_{10}$S$_4$Se$_4$H$_8$] molecules
(Fig.~\ref{figStruct}). The bulk magnetic properties of the system are
dominated by the paramagnetic 3\emph{d} electron spin moments of Mn$^{2+}$ ions
($S_d=5/2, g\approx 2$) located in the anion layers \cite{r12,VyaselPRB2011}.
The (BETS)$_2^+$ dimers accommodating one hole per dimer, form an effectively
half-filled quasi-two-dimensional $\pi$-electron conduction band
\cite{ZverevPT}. The compound undergoes a metal-insulator (MI) transition at
cooling below $T_{\rm{MI}}\sim23\,$K at ambient pressure \cite{r12}. The
transition temperature rapidly decreases at applying pressure, $P$, so that at
$P\geq1\,$kbar the system is metallic in the whole temperature range
\cite{ZverevPT}. Moreover, it goes superconducting with maximum $T_c=5.8\,$K at
$P=0.6$\,-1.0\,kbar.

Much effort has been done recently to understand the influence of paramagnetic
$d$-metal ions of the anion layers on the transport properties of
quasi-two-dimensional systems \cite{KobaJACS118, KobaChemRev}. It has been
undoubtedly established now that the field-induced superconductivity in
$\lambda$-(BETS)$_2$FeCl$_4$ and $\kappa$-(BETS)$_2$FeBr$_4$ arises when the
external magnetic field compensates the exchange field created by localized
3$d$ moments of Fe$^{3+}$ on the spins of itinerant $\pi$-electrons
\cite{UjiJPSJ75}. The role of $\pi$-$d$ interactions in driving the MI
transition observed in the systems with magnetic anions is not as clear. For
example, at ambient pressure and zero field, non-magnetic
(BETS)$_2$Ga(Cl,Br)$_4$ systems are metallic down to lowest temperatures,
whereas the iron-containing $\kappa$-(BETS)$_2$Fe(Cl,Br)$_4$ are metallic below
the antiferromagnetic (AF) transition, and $\lambda$-(BETS)$_2$FeCl$_4$ is a
uniaxial N\'{e}el-type AF insulator below $T_N=T_{\rm{MI}}=8\,$K
\cite{KobaChemRev}. It has been suggested \cite{BrossFe} that in the
abovementioned Fe-containing systems the AF state is initiated within the
system of 3\emph{d} electron spin moments of Fe$^{3+}$ ions, which in the case
of $\lambda$-(BETS)$_2$FeCl$_4$ drives the MI transition as well. This
reasonable assumption has been doubted however in later communications
\cite{AkibaSH,BrooksMoess}.

Despite the similarity of the title compound to the listed systems with FeX$_4$
(X=Cl, Br) anions, including the presence of the 3\emph{d} ions in the anion
layer and the MI transition, its phase diagram is different. First, at ambient
pressure the system remains insulating in the external magnetic field up to at
least 15\,T with no sign of the field-induced superconductivity. Secondly, its
bulk dc magnetization does not demonstrate a pronounced AF transition
\cite{VyaselPRB2011}. A feasible reason for the latter is considerably
different structure of its anion layer. In the iron-based compounds, it is a
rectangular network of discrete paramagnetic units FeX$_4^-$, so that the
exchange coupling between Fe$^{3+}$ ions is mediated by $\pi$-electrons of
cation layers \cite{BrossFe}. In BETS-Mn, the Mn atoms are arranged in
isosceles triangles, a geometry that is frustrating for an antiparallel spin
order. Besides, the $\pi$-electrons are not necessarily involved into the
exchange interaction between Mn$^{2+}$ ions linked within the polymeric anion
layer via dicyanamide bridges \cite{r12}.

In fact, below the MI transition the bulk static susceptibility of BETS-Mn
deviates slightly from the Curie-Weiss law obeyed accurately above
$T_{\rm{MI}}$. Gradually decreasing compared to the Curie-Weiss value,
$\chi_{\rm{CW}}(T)$, the susceptibility becomes 80-85\% of $\chi_{\rm{CW}}$ at
$T=2\,$K \cite{VyaselPRB2011}. Combined with the results of $^1$H NMR
\cite{VyaselJETP} this infers some disordered tilt of the static component of
Mn$^{2+}$ moments from the external field direction. This could possibly result
from the trend of the Mn$^{2+}$ system towards AF order, frustrated
geometrically by the triangular arrangement of Mn in the anion layer. This
magnetic effect occurs just below the MI transition temperature indicating the
existence of the coupling between $d$ electrons of Mn$^{2+}$ and the conduction
$\pi$-electrons. However, the small size of this effect and its gradual
character (compared to the steep resistivity growth below $T_{\rm{MI}}$) makes
it unlikely to be a driving force for the MI transition. Besides, the magnetic
torque measurements have given a hint of a magnetic order within
$\pi$-electrons below the MI transition \cite{VyaselPRB2011}. This favors a MI
transition scenario more typical for quasi-two-dimensional conductors with
half-filled band which involves strong correlations within the $\pi$-electrons
leading to a Mott instability \cite{KanodaNPhys5}, as it has been suggested in
\cite{ZverevPT, VyaselPRB2011}.

An insight into the properties of the $\pi$-electron spin system is important
for illuminating this issue. Invisible in the bulk magnetization measurements
due to the masking effect from Mn$^{2+}$ moments, the $\pi$-spin system can be
effectively probed using NMR on the ``central'' carbons of the BETS molecule,
Fig.~\ref{figStruct} \cite{KlutzApMagRes2}. In this paper we report the results
of $^{13}$C NMR experiments on BETS-Mn performed at ambient pressure and under
an external pressure sufficient to suppress the MI transition.

\begin{figure}[p]
\includegraphics[width=0.8\linewidth]{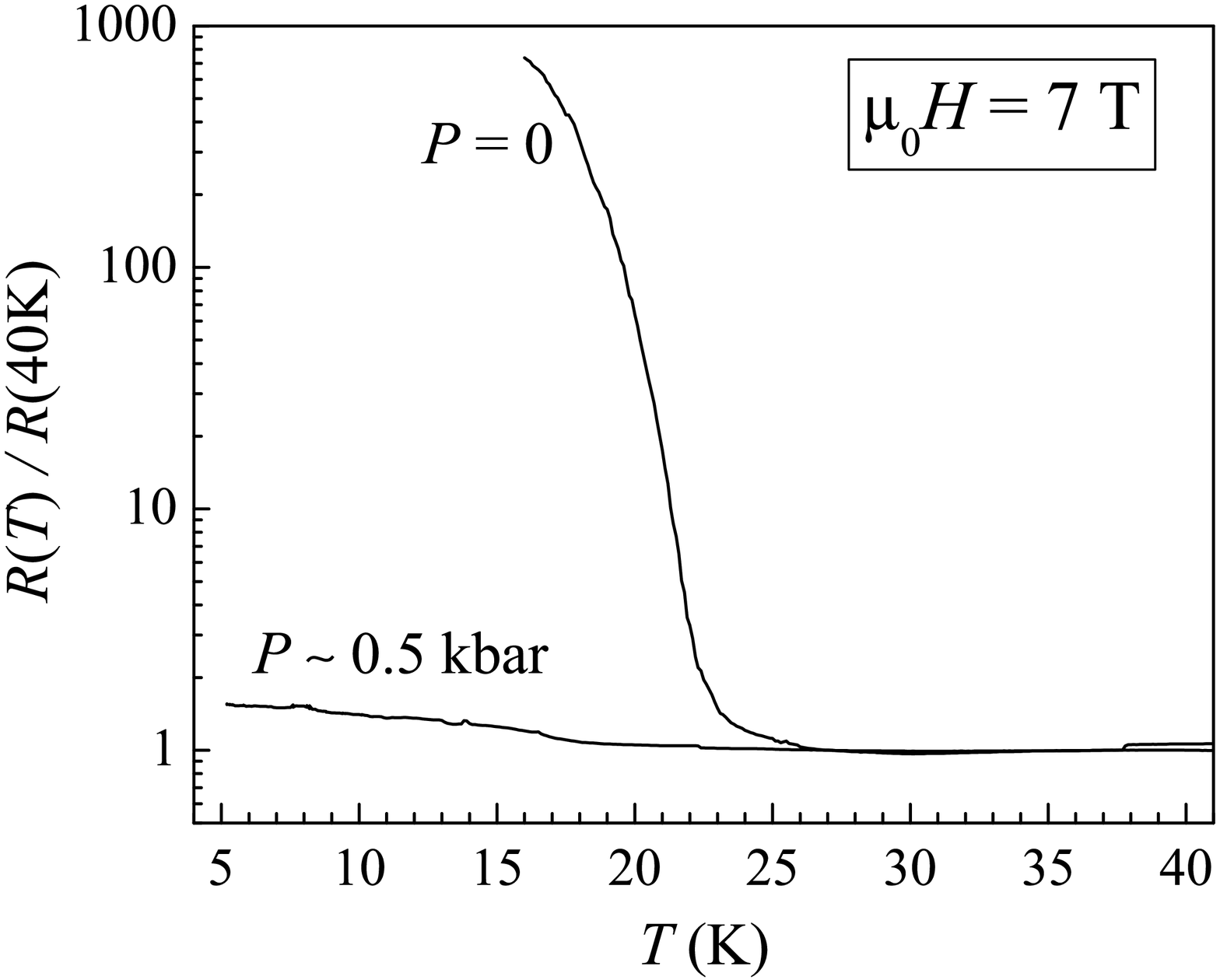}
\caption{Temperature dependences of the interplane resistivity of BETS-Mn
crystal at ambient pressure and under a pressure of $\sim 0.5\,$kbar, measured
in the external field $\mu_0 H=7\,$T. \label{figRvsT}}
\end{figure}

The sample was a single crystal of BETS-Mn with BETS molecules containing (at
least) 99\% of $^{13}$C isotope in the positions of the central carbons,
Fig.~\ref{figStruct}. The sample size was $a^\ast\times b\times c\sim
0.05\times 3\times 3\,$mm$^3$. BETS-Mn crystals were synthesized by
electrochemical oxidation of the $^{13}$C-labeled BETS in the presence of the
complex salt Ph$_4$PMn[N(CN)$_2$]$_3$ as electrolyte according to the procedure
described in Ref.~\onlinecite{KushchJSSC184}. The sample was suspended inside a
$5\times 5\times 0.5\,$mm$^3$ NMR rf coil on four 10\,$\mu$m-thick platinum
leads attached to the sample to provide the resistivity measurements across the
conducting $(b,c)$ plane, and sealed inside a PTFE can. To impose the external
pressure to the sample, the can was filled with GKZh-136 silicone liquid which
is known to generate at low temperatures a pressure of several hundred bar due
to the difference in thermal contractions \cite{GKZh}. To testify the pressure
generated inside the sample, the resistance of the sample with and without
GKZh-136 was measured in the external field $\mu_0 H=7\,$T in the same cooling
cycles with NMR measurements. Figure~\ref{figRvsT} demonstrates essential
suppression of the MI transition in the sample submerged in GKZh-136 implying a
pressure of at least 0.5\,kbar generated in the sample \cite{ZverevPT}.

NMR was measured in field $\mu_0 H=7\,$T aligned perpendicular to [0$\bar{1}$1]
diagonal at angle 45$^\circ$ with $a^\ast$ direction. The spectra were acquired
using a standard spin-echo sequence with $\pi$-pulse length $\leq2.5\,\mu$s. To
cover broad spectra, Fourier-transforms of the acquired spin-echoes were
collected at intervals $\leq$150~kHz and summed up.

\begin{figure}[p]
\includegraphics[width=0.9\linewidth]{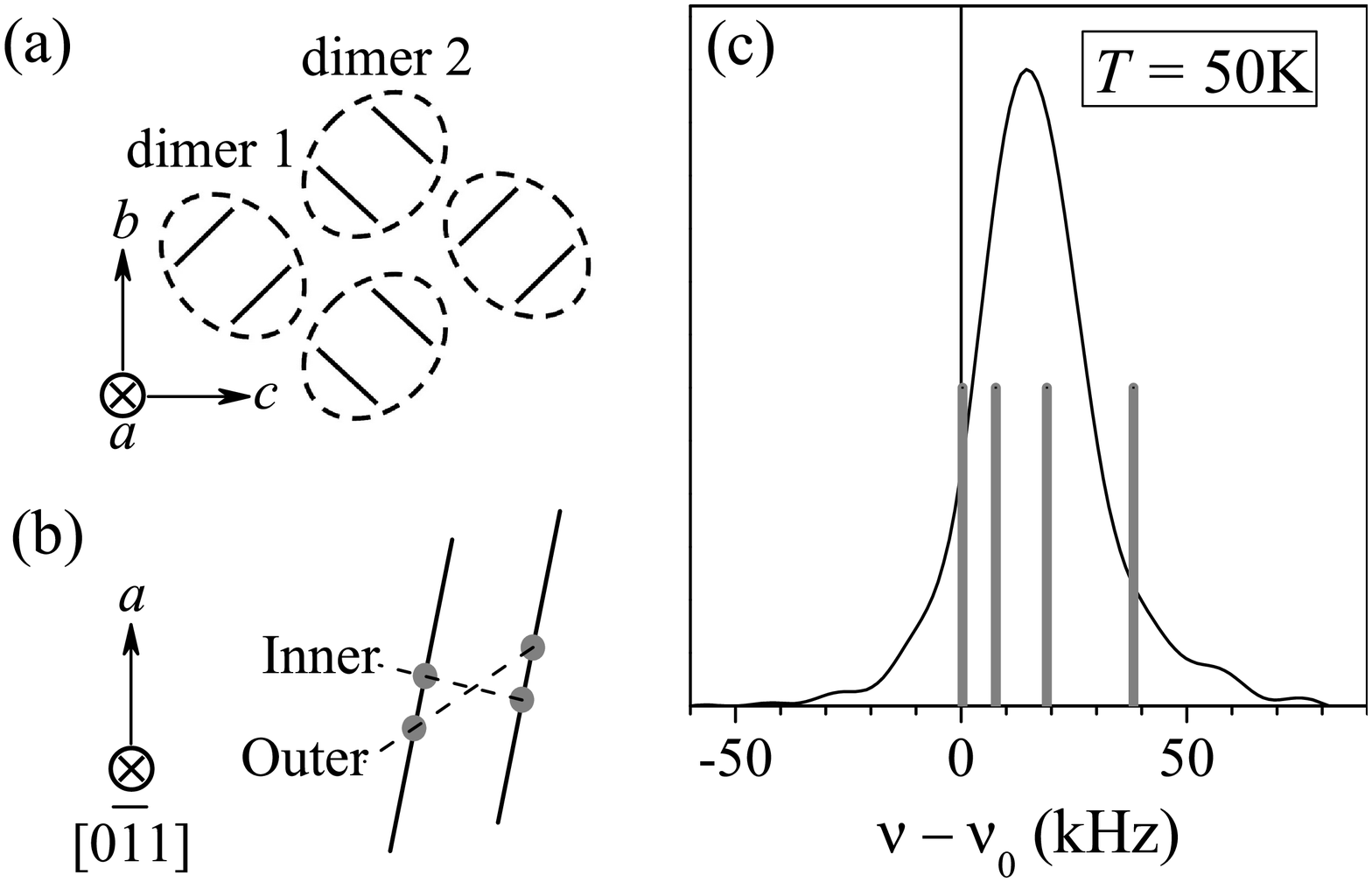}
\caption{(a) Schematic top view of the conducting BETS layer. (BETS)$_2$ dimers
are represented by the pairs of lines. (b) Schematic side view of a BETS dimer
with site definition of the central carbons. (c) $^{13}$C NMR spectrum measured
at $T$\,=\,50\,K in the external field $\mu_0 H$\,=\,7\,T aligned perpendicular
to $[0\bar{1}1]$ at 45$^\circ$ to $\vec{a}^\ast$. The spectrum is shown with
respect to $\nu_0=74.946\,$MHz. Grey vertical lines indicate positions of the
resonance peaks calculated with the shift tensor of
$\kappa$-(ET)$_2$Cu[N(CN)$_2$]Br. \label{figMetSpec}}
\end{figure}

For the experimental geometry of our choice, $H\bot [0\bar{1}1]$, $\angle
(H,a^\ast)=45^\circ$, the $^{13}$C NMR spectrum measured in the metallic state
of the title compound is a single resonance line, as exemplified in
Figure~\ref{figMetSpec}(c). In theory, for the given field orientation one
expects a BETS-Mn crystal to produce four $^{13}$C NMR peaks
\cite{DeSotoPRB52}. They arise from the two magnetically different orientations
of the BETS dimers in the unit cell, as shown in Fig.~\ref{figMetSpec}(a), and
nonequivalent (`inner' and `outer') central carbon sites within the dimers,
Fig.~\ref{figMetSpec}(b). The BETS molecules within the dimer are
inversion-symmetric to each other, thus magnetically equivalent. The dipolar
interaction between the spins of the central $^{13}$C, which in general
provides another factor of two to the number of peaks, is nearly zero for this
field orientation.

In order to calculate the positions of the four peaks one needs to know the
values of the $^{13}$C shift tensors for the `inner' and `outer' carbon sites
in BETS-Mn. The total shift tensor is composed of the chemical (orbital),
Knight (spin), and ``dipolar'' shift tensors. The dipolar contribution
resulting from the dipolar fields induced by Mn$^{2+}$ moments can be
calculated following Ref.~\onlinecite{VyaselJETP}, using the measured dc
magnetization values \cite{VyaselPRB2011}. The former two contributions, which
characterize, respectively, the intrinsic fields within the BETS molecule
itself and the fields induced by the spins of conduction electrons, are
currently unknown. Assuming that the tensors are similar to those in the
ET-based compounds (ET is isostructural to BETS but with sulphur on Se sites),
we use the values obtained for $\kappa$-(ET)$_2$Cu[N(CN)$_2$]Br
\cite{DeSotoPRB52} to calculate the $^{13}$C NMR peak positions. The resulting
peak positions (including the dipolar shift) are indicated in
Fig.~\ref{figMetSpec}(c) by grey vertical lines. One can see that the resonance
line observed in the experiment covers fairly well the range of the calculated
peak positions, indicating that the $^{13}$C shift tensor in BETS-Mn is not
much different from that in $\kappa$-(ET)$_2$Cu[N(CN)$_2$]Br.

Evidently, the peaks from individual carbon sites in the metallic state of
BETS-Mn are vastly broadened and merged into a single line. An excessive
broadening of $^{13}$C NMR peaks has been noticed in
$\kappa$-(ET)$_2$Cu[N(CN)$_2$]Br below $T$\,=\,150\,K, and a number of reasons
has been recruited to explain it, including spacial variation of the
$\pi$-electron spin density due to precursors to Anderson localization or a
spin-density wave \cite{DeSotoPRB52}. Minor sample imperfection, which cannot
be ruled out, may also contribute to the linewidth because each of the three
contributions to the total shift mentioned above are highly anisotropic, while
the dipolar shift is also sensitive to atomic displacements \cite{VyaselJETP}.
More detailed analysis of possible broadening mechanisms in the metallic state
is beyond the scope of this communication.

\begin{figure}[p]
\includegraphics[width=1.0\linewidth]{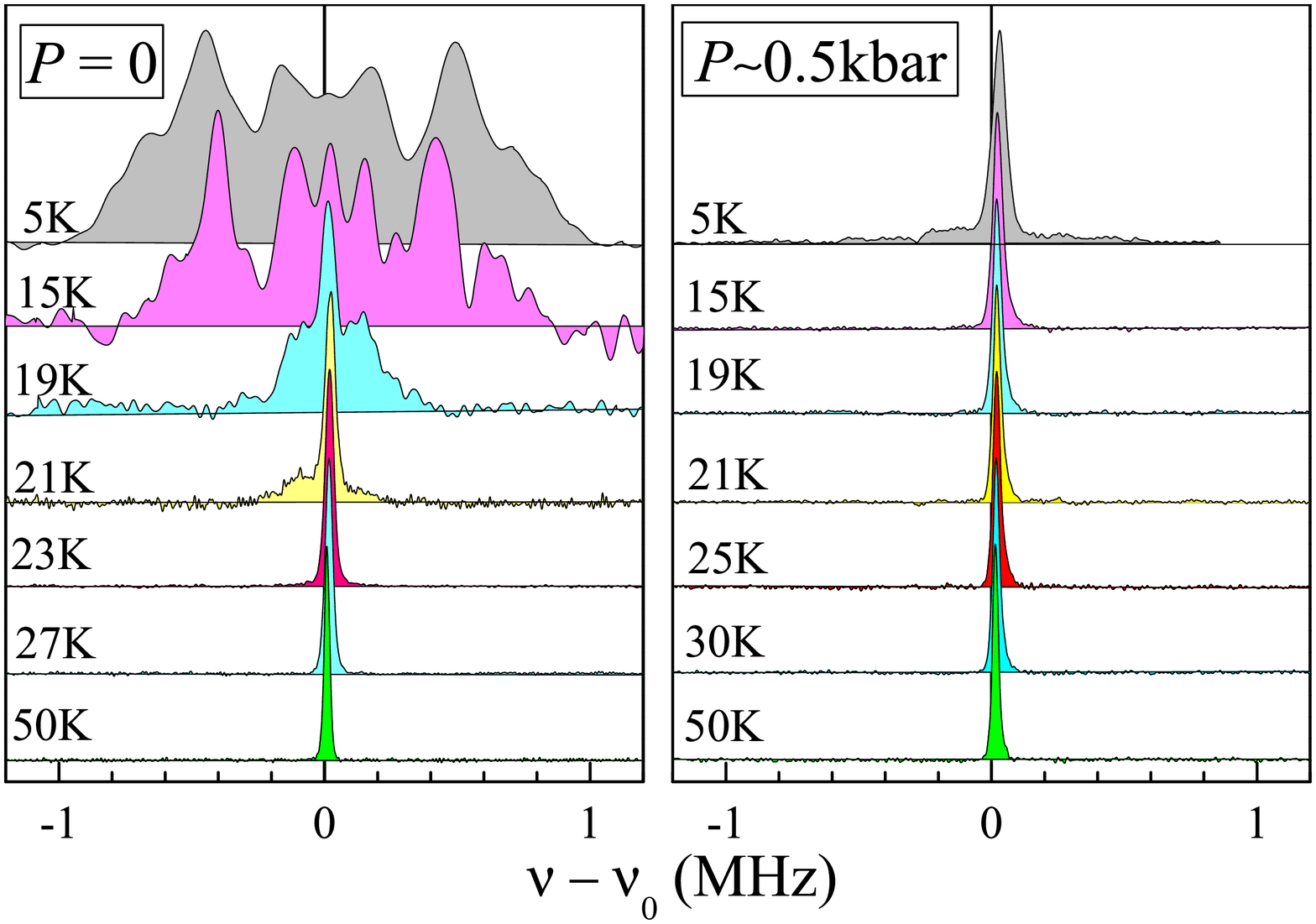}
\caption{(Color online) $^{13}$C NMR spectra measured in BETS-Mn at several
temperatures at ambient pressure, $P=0$ (left) and in GKZh-136, $P\sim
0.5\,$kbar, (right). \label{Specs}}
\end{figure}

The shape of the $^{13}$C NMR spectrum changes rapidly as the system enters the
insulating state. The left panel in Figure~\ref{Specs} shows $^{13}$C NMR
spectra measured in BETS-Mn at temperatures from 5 to 50\,K at ambient
pressure. The single peak characteristic of the spectrum in the metallic state
above $T=23$\,K$\approx T_{\rm{MI}}$, develops below this temperature into a
broad symmetric pattern counting 5 peaks at least. At $T=5\,$K the spectrum
spans a range of nearly $\pm 1\,$MHz which is huge compared to the spectrum in
the metallic state. This cannot result from the dipolar fields created by
Mn$^{2+}$: calculations show that fully polarized Mn$^{2+}$ can provide a
dipolar shift ranging from $-12.5$ to $-19\,$kHz (depending on the carbon site)
in a 7\,T field. Therefore, the spectrum below $T_{\rm{MI}}$ evidences an
enhancement of the hyperfine field experienced by the $^{13}$C nuclear spin due
to localization of the $\pi$-electron spins on the dimers of the BETS
molecules. Moreover, several pronounced peaks are visible in the
low-temperature spectrum, which infers a commensurate order of the localized
spins. Finally, the symmetric shape of the spectrum indicates the staggered
order, because antiparallel components of the staggered electron spins
($\hat{\textrm{\textbf{S}}}_i=-\hat{\textrm{\textbf{S}}}_j$) produce opposite
local fields at carbon sites \textit{i} and \textit{j}:
$h_{i}=\hat{\textrm{\textbf{I}}}_i\cdot \textsf{A}\cdot
\hat{\textrm{\textbf{S}}}_i=-(\hat{\textrm{\textbf{I}}}_j\cdot \textsf{A}\cdot
\hat{\textrm{\textbf{S}}}_j)=-h_{j}$, where $\hat{\textrm{\textbf{I}}}$ and
$\hat{\textrm{\textbf{S}}}$ are the nuclear and electron spin operators,
respectively, and $\textsf{A}$ is the hyperfine tensor. In turn, this signifies
the AF interaction between the localized spins. The staggered component of the
spin magnetization should lie somewhere in the plane perpendicular to the
magnetic field, since the anisotropic exchange term is usually much smaller
than the external field $\mu_0 H$\,=\,7\,T.

These signs of the commensurate staggered spin order are only present in the
insulating state of BETS-Mn. The right panel in Fig.~\ref{Specs} shows the
spectra for the sample submerged in GKZh. According to the resistance
measurements (Fig.~\ref{figRvsT}), the MI transition in this case is
substantially suppressed due to the pressure ($P\sim 0.5\,$kbar) imposed to the
sample by thermal contraction of GKZh. The narrow-peak spectrum characterizing
the metallic state, obviously survives way below 23\,K. Only at $T=5\,$K the
broad features show up, reminiscent of the ambient-pressure spectrum at
$T=21\,$K.

Therefore, the discovered staggered order of the localized spins is obviously
related to the insulating state of BETS-Mn (cf. Fig.~\ref{figRvsT} and
Fig.~\ref{Specs}). First, it arises at the MI transition temperature. Secondly,
it is suppressed by the external pressure in the same way as the MI transition.
It is evident therefore that the staggered order occurs within the spin system
of $\pi$ electrons which localize on the dimers of the BETS molecules below
$T_{\rm{MI}}$. This finding provides a strong support in favor of an AF
Mott-insulating state as a ground state of BETS-Mn.


The frequency range of the $^{13}$C spectrum in BETS-Mn at $T=5\,$K measured at
ambient pressure (left panel in Fig.~\ref{Specs}) is of the same order of
magnitude ($\pm 1\,$MHz) as in the AF Mott state of
$\kappa$-(ET)$_2$Cu[N(CN)$_2$]Cl \cite{SmithPRB68} and
$\beta^\prime$-(ET)$_2$ICl$_2$ \cite{EtoPRB81}. A magnitude of the electron
spin magnetization of 0.5\,$\mu_{\rm{B}}$ and 1\,$\mu_{\rm{B}}$ per dimer,
respectively, has been reported for these two compounds. It should be within
the same range here, since the hyperfine tensor in BETS-Mn is expected to be
similar, as the measurements in the metallic phase suggest. It should be noted
however that the staggered spin structure discovered in the present compound
does not necessarily signify the conventional N\'{e}el AF state but can be a
field-induced effect, as it occurs in $\kappa$-(ET)$_2$Cu[N(CN)$_2$]Cl above
$T_N$ \cite{KanodaPRB78}.

The staggered spin structure of the localized $\pi$ electrons can be
responsible for the magnetic effect observed in the 3\emph{d} Mn$^{2+}$
electron spin system. The behavior of the Mn$^{2+}$ system below $T_{\rm{MI}}$
examined by the static magnetization and $^1$H NMR measurements
\cite{VyaselJETP,VyaselPRB2011} infers some disordered tilt of the static
component of Mn$^{2+}$ moments from the external field direction. This can be
interpreted as a trend of the Mn$^{2+}$ system towards AF order, frustrated
geometrically by the triangular arrangement of Mn in the anion layer. However,
the susceptibility of BETS-Mn determined by the Mn$^{2+}$ spin system, depends
on temperature as $(T+\theta)^{-1}$ with $\theta\approx$\,5.5\,K
\cite{VyaselPRB2011}, which implies a paramagnetic state of the Mn$^{2+}$
subsystem down to $T\sim\theta$. In fact, it deviates from the paramagnetic
behavior at much higher temperature $T_{\rm{MI}}\sim$\,23\,K. One can assume
therefore that it is the ordered $\pi$ spin system that affects the
paramagnetic state of Mn$^{2+}$ spins via the $\pi$-\emph{d} interactions. The
disorder in the frustrated Mn$^{2+}$ system, in turn, introduces some disorder
in the $\pi$-spin structure. Comparing $^{13}$C NMR spectra taken at 5 and
15\,K at ambient pressure (left panel in Fig.~\ref{Specs}), one can notice that
the positions of the resonance peaks are nearly the same, indicating a nearly
constant magnitude of the electron spin magnetization, while the peak widths
are much bigger at $T$\,=\,5\,K. It is possible that the disordered Mn$^{2+}$
spins whose magnetization grows in this temperature region by a factor of two
\cite{VyaselPRB2011}, induce some disorder in the orientations of the
$\pi$-spins. In other words, the trend of the Mn$^{2+}$ spins to follow the AF
arrangement of the $\pi$-spins is frustrated by the triangular network of Mn,
which in turn leads to some disorder of the $\pi$-spins as well.

Indeed, determining the spin structure of the localized $\pi$ electrons would
be of great importance. To do this one needs to derive from the NMR data the
hyperfine tensors for the central carbons in BETS-Mn. This requires the value
of the $\pi$ spin susceptibility which, to our knowledge, has never been
reported so far. An appropriate technique to measure this value would be ESR,
provided that the resonance peaks from Mn$^{2+}$ and $\pi$ electron spins are
resolved. Besides, ESR could give information about the low-field spin
structure of the localized $\pi$ electrons, such as the orientation of the
easy-axis (if exists).

In conclusion, we performed $^{13}$C NMR measurements on the central carbons of
the BETS molecules in BETS-Mn with and without the external pressure to
retrieve the spin properties of the conduction $\pi$ electrons. We found that
the transition of the system into the insulating state is accompanied by
ordering of the $\pi$ spins into a long-range staggered structure. This
supports the assumption that the MI transition is due to Mott instability
resulting from strong correlations within the half-filled band conduction
electron system. The staggered $\pi$-spin structure is suggested to be
responsible for the changes in the Mn$^{2+}$ spin system observed in the static
magnetization and $^1$H NMR experiments.

The authors thank R.~Kato, H.~M.~Yamamoto M.~Kobayashi for providing
$^{13}$C-labeled BETS. We gratefully acknowledge fruitful suggestions from
S.~E.~Brown, the assistance in questions of crystallography from S.~V.~Simonov,
and technical support from N.~A.~Belov. This work was supported by the RFBR
grants 10-02-01202 and 11-02-91338-DFG (DFG grant Bi 340/3-1) and the Russian
State Contract 14.740.11.0911.

\end{document}